\documentclass[twocolumn,aps,superscriptaddress,showpacs,amsmath,amssymb,prl]{revtex4-1}
\usepackage{graphicx}
\usepackage{dcolumn}
\usepackage[english]{babel}
\usepackage{amsmath,amssymb,amsfonts}
\usepackage{graphicx}
\usepackage{bm}

\usepackage{array}
\usepackage{color}
\usepackage{multirow}
\usepackage[normalem]{ulem}
\usepackage[caption=false]{subfig}
\usepackage{siunitx}
\usepackage{soul}
\usepackage{upgreek}
\usepackage{lineno}

\newcommand{\MBS}{MnBi$_2$Se$_4$}
\newcommand{\MBT}{MnBi$_2$Te$_4$}
\newcommand{\MST}{MnSb$_2$Te$_4$}

\newcommand{\BS}{Bi$_2$Se$_3$}

\newcommand{\PreserveBackslash}[1]{\let\temp=\\#1\let\\=\temp}
\let\PBS=\PreserveBackslash

\begin{document}


\title{Unique thickness-dependent properties of the van der Waals interlayer antiferromagnet \MBT\, films}

\author{M.\,M. Otrokov}
\email{mikhail.otrokov@gmail.com}
\affiliation{Centro de F\'{i}sica de Materiales (CFM-MPC), Centro Mixto CSIC-UPV/EHU,  20018 Donostia-San Sebasti\'{a}n, Basque Country, Spain}
\affiliation{Tomsk State University, 634050 Tomsk, Russia}

\author{I.\,P. Rusinov}
\affiliation{Tomsk State University, 634050 Tomsk, Russia}
\affiliation{Saint Petersburg State University, 198504 Saint Petersburg, Russia}

\author{M.\, Blanco-Rey}
\affiliation{Departamento de F\'{\i}sica de Materiales UPV/EHU, 20080 Donostia-San Sebasti\'{a}n, Basque Country, Spain}
\affiliation{Donostia International Physics Center (DIPC), 20018 Donostia-San Sebasti\'{a}n, Basque Country, Spain}

\author{M.\, Hoffmann}
\affiliation{Institut f\"ur Theoretische Physik, Johannes Kepler Universit\"at, A 4040 Linz, Austria}

\author{A.\,Yu. Vyazovskaya}
\affiliation{Tomsk State University, 634050 Tomsk, Russia}
\affiliation{Saint Petersburg State University, 198504 Saint Petersburg, Russia}

\author{S.\,V.~Eremeev}
\affiliation{Institute of Strength Physics and Materials Science, Russian Academy of Sciences, 634021 Tomsk, Russia}
\affiliation{Tomsk State University, 634050 Tomsk, Russia}
\affiliation{Saint Petersburg State University, 198504 Saint Petersburg, Russia}

\author{A.~Ernst}
\affiliation{Institut f\"ur Theoretische Physik, Johannes Kepler Universit\"at, A 4040 Linz, Austria}
\affiliation{Max-Planck-Institut f\"ur Mikrostrukturphysik, Weinberg 2, D-06120 Halle, Germany}

\author{P.\,M. Echenique}
\affiliation{Centro de F\'{i}sica de Materiales (CFM-MPC), Centro Mixto CSIC-UPV/EHU,  20018 Donostia-San Sebasti\'{a}n, Basque Country, Spain}
\affiliation{Departamento de F\'{\i}sica de Materiales UPV/EHU, 20080 Donostia-San Sebasti\'{a}n, Basque Country, Spain}
\affiliation{Donostia International Physics Center (DIPC), 20018 Donostia-San Sebasti\'{a}n, Basque Country, Spain}

\author{A. Arnau}
\affiliation{Centro de F\'{i}sica de Materiales (CFM-MPC), Centro Mixto CSIC-UPV/EHU,  20018 Donostia-San Sebasti\'{a}n, Basque Country, Spain}
\affiliation{Departamento de F\'{\i}sica de Materiales UPV/EHU, 20080 Donostia-San Sebasti\'{a}n, Basque Country, Spain}
\affiliation{Donostia International Physics Center (DIPC), 20018 Donostia-San Sebasti\'{a}n, Basque Country, Spain}

\author{E.\,V. Chulkov}
\affiliation{Centro de F\'{i}sica de Materiales (CFM-MPC), Centro Mixto CSIC-UPV/EHU,  20018 Donostia-San Sebasti\'{a}n, Basque Country, Spain}
\affiliation{Departamento de F\'{\i}sica de Materiales UPV/EHU, 20080 Donostia-San Sebasti\'{a}n, Basque Country, Spain}
\affiliation{Donostia International Physics Center (DIPC), 20018 Donostia-San Sebasti\'{a}n, Basque Country, Spain}
\affiliation{Saint Petersburg State University, 198504 Saint Petersburg, Russia}

\date{\today}

\begin{abstract}
Using density functional theory and Monte Carlo calculations, we study the thickness dependence 
of the magnetic and electronic properties of a van der Waals interlayer antiferromagnet in the 
two-dimensional limit. Considering \MBT\, as a model material, we find it to demonstrate a 
remarkable set of thickness-dependent magnetic and topological transitions. 
While a single septuple layer block of \MBT\, is a topologically
trivial ferromagnet, the thicker films made of an odd (even) number of blocks are
uncompensated (compensated) interlayer antiferromagnets, which show wide bandgap quantum 
anomalous Hall (zero plateau quantum anomalous Hall) states.
Thus, \MBT\, is the first stoichiometric material predicted to realize the zero plateau 
quantum anomalous Hall state intrinsically. This state has been theoretically shown to host the exotic axion insulator phase.
\end{abstract}

\maketitle

After the isolation of graphene, the field of two-dimensional (2D) van der Waals (vdW) materials 
has experienced an explosive growth and new families of 2D systems and block-layered bulk materials, 
such as tetradymite-like topological insulators (TIs) \cite{Hasan2010, Eremeev.ncomms2012}, 
transition metal dichalcogenides \cite{Xu.nphys2014}, and others \cite{Ishizaka.nmat2011, 
Li.nnano2014, Banerjee.nmat2016} have been discovered. 
The remarkable electronic properties, along with 
the possibility of their tuning via thickness control, doping, intercalation, proximity effects, 
etc., make the layered vdW materials attractive from both practical and fundamental points of view. The relative 
simplicity of fabrication with a number of techniques has greatly facilitated a comprehensive 
study of these systems. However, the important step towards magnetic functionalization of the 
inherently nonmagnetic layered vdW materials and a controllable fabrication of the resulting hybrid 
systems has proven challenging. 
Therefore, 
aiming at exploring magnetism of layered vdW materials in the 2D limit, new possibilities have been considered. One of them is the ultrathin laminae exfoliation from intrinsically ferromagnetic (FM) vdW crystals, such as Cr$_2$Ge$_2$Te$_6$ and CrI$_3$, whose magnetic behaviour  has been studied down to a few layers thickness \cite{Gong.nat2017,Huang.nat2017}. 
An alternative fabrication strategy is epitaxial growth \cite{Hirahara.nl2017, Hagmann.njp2017}. 
With this technique, a 2D FM septuple layer (SL) block of MnBi$_2$Se$_4$ has been grown on top of the \BS\, TI surface  \cite{Hirahara.nl2017}.
Similar systems have been theoretically proposed as a promising platform for achieving the quantized anomalous Hall (QAH) and magnetoelectric effects at elevated temperatures \cite{Otrokov.jetpl2017, Otrokov.2dmat2017}. Later, epitaxial growth of the Bi$_2$Se$_3$/MnBi$_2$Se$_4$ multilayer heterostructure has been reported \cite{Hagmann.njp2017}.

The field of 2D vdW magnets is in its infancy and many more materials with new properties are to be
explored. In particular, vdW \textit{antiferromagnets} are expected to be of great interest. Indeed, 
recently it has been reported that the 
layered vdW compound \MBT\, is the first ever antiferromagnetic (AFM) TI \cite{Otrokov.arxiv2018}. This state of matter 
is predicted to give rise to exotic phenomena such as quantized magnetoelectric 
effect \cite{Mong.prb2010}, axion 
electrodynamics \cite{Li.nphys2010}, and Majorana hinge modes \cite{Peng.arxiv2018}. 
Moreover, the combination of magnetism with spin-orbit coupling, along with strong thickness dependence
of electronic structure in the 2D limit suggest that vdW compounds like $\mathrm{MnBi_2Te(Se)_4}$ and \MST\,  \cite{Eremeev.jac2017} might be attractive for both fundamental and applied research. 
Finally, a novel type of energetically stable and universal interface between (A)FM and topological 
insulators has been proposed recently \cite{Eremeev.nl2018}. At such an interface, the film of magnetic material, that
does not show vdW bonding intrinsically, turns out to be vdW-coupled to a TI as a result of immersion 
below the surface of the latter. 
Incidentally, the axion insulator state could also be achieved in such heterostructures \cite{Hou.arxiv2018}. 
These and other AFM systems appear to be interesting candidates to couple the emerging fields of AFM spintronics 
\cite{Baltz.rmp2018} and layered vdW materials \cite{Gong.nat2017,Huang.nat2017}. 

\begin{figure}[!bth]
\begin{center}
\includegraphics[width=0.400\textwidth]{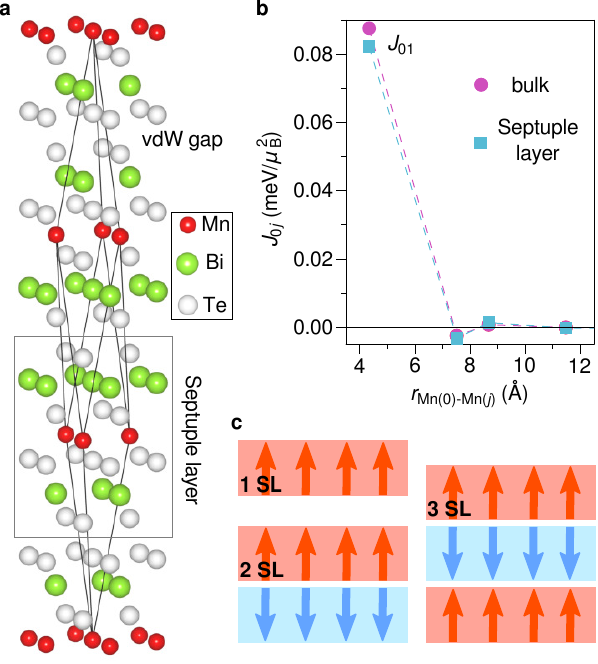}
\end{center}
\caption{(a) Atomic structure of the $R\bar 3m$-group bulk \MBT\, 
with red, green, and white balls showing Mn, Bi, and Te atoms, respectively. 
The paramagnetic rhombohedral unit cell is shown by black lines.
(b) Calculated exchange constants $J_{0j}$ for the 
intralayer pair interactions as a function of the distance 
$r_{\text{Mn}(0)-\text{Mn}(j)}$ for the bulk (circles) and 
free-standing SL (squares). (c) 1-, 2-, and 3-SL-thick \MBT\, films, 
showing an FM, compensated AFM, and uncompensated AFM orders, respectively.}
\label{fig1}
\end{figure}

Here, using state-of-the-art \emph{ab initio} techniques and the Monte Carlo method, we study the magnetic, 
electronic and topological properties of the layered vdW AFM TI compound MnBi$_2$Te$_4$ 
in the 2D limit. We find a unique set of thickness-dependent magnetic and topological transitions, 
which drive the \MBT\, thin films through FM and (un)compensated AFM phases (see Fig.~\ref{fig1}), as well as QAH and zero plateau QAH states. 

The electronic structure calculations
were carried out within density functional theory using the
projector augmented-wave method \cite{Blochl.prb1994} (VASP code 
\cite{vasp1,vasp2}). The exchange-correlation energy was treated using the
generalized gradient approximation \cite{Perdew.prl1996}. The
Hamiltonian contained scalar relativistic corrections and the
spin-orbit coupling was taken into account \cite{Koelling.jpc1977}. 
To describe the vdW
interactions we used the DFT-D3 approach \cite{Grimme.jcp2010,
Grimme.jcc2011}. 
The Mn $3d$-states were
treated employing the GGA$+U$ approach \cite{Anisimov1991, Dudarev.prb1998}. 
The Heisenberg exchange coupling constants $J_{ij}$ were
computed \emph{ab initio} using the full-potential linearized augmented plane waves
method \cite{bib:wimmer81} (\textsc{Fleur} code \cite{bib:fleur}).
The magnetic critical temperatures were determined using Monte Carlo simulations based on a
classical Heisenberg Hamiltonian parametrised with  the magnetic anisotropy energies (MAEs) and $J_{ij}$ constants obtained from \textit{ab
initio} calculations. The Chern numbers were 
determined using Z2Pack \cite{Soluyanov.prb2011, Gresch.prb2017}
and \emph{ab-initio}-based tight-binding calculations 
~\cite{marzari_vanderbilt1997,Mostofi2008}.
The edge electronic band structure was calculated within the semi-infinite medium
Green function approach. More details on methods can be found in the 
Supplementary Note I.

\MBT\, is built of SL blocks stacked one on
top of another along the [0001] direction and held together by 
vdW forces \cite{Lee.cec2013, Otrokov.arxiv2018} (Fig.~\ref{fig1}a). 
As far as its magnetic order is
concerned, it appears to be an interlayer antiferromagnet, where   
the FM Mn layers of neighboring blocks are coupled
antiparallel to each other \cite{Eremeev.jac2017,
Otrokov.arxiv2018}. Although the recently synthesized
single crystals show some degree of statistical Mn/Bi disorder, our
\emph{ab initio} calculations indicate that such an intermixing is less
favorable than the ideal structure (Supplementary Note II). 
Therefore, in what follows we consider the ordered structure of \MBT.

We start with the magnetic characterization of a single free-standing \MBT\, SL.
The exchange coupling constants $J_{0j}$, calculated as a function of the Mn-Mn 
distance $r_{\text{Mn}(0)-\text{Mn}(j)}$, show the same trend as in bulk \MBT\ (Fig.~\ref{fig1}b).
Namely, that the FM interaction between first nearest neighbors, $J_{01}\simeq 
\SI{0.08}{\meV}/\mu_\mathrm{B}^2$, strongly dominates over all other. Thus, there 
is a stable tendency towards the FM ground state in \MBT\, SL, as confirmed by 
the total-energy calculations, which show a preference of the FM state over the 
120$^\circ$ AFM state by 14.77 meV per Mn pair (Table~\ref{tab:mag}). This 
result is consistent with those of Refs.~\cite{Otrokov.arxiv2018, Gong.arxiv2018}, 
where it has been experimentally shown that each \MBT\, SL orders ferromagnetically. 
An FM ground state has also been predicted for a single \MBT\, SL placed on 
different tetradymite-like TI substrates \cite{Otrokov.jetpl2017, Otrokov.2dmat2017}. 
Therefore, we conclude that the intralayer FM order is not sensitive to the 
thickness of the \MBT\, film, as well as to the formation of the vdW interface with other block-layered compounds.

{\renewcommand{\arraystretch}{1.00}%
\begin{table}[!bth]
\caption{
Thickness dependence of the MnBi$_2$Te$_4$ films magnetism.
$\Delta E_\text{A/F}= E_\text{AFM} - E_\text{FM}$ is the total energy difference
of the AFM and FM states, where AFM refers to the intralayer 120$^\circ$
state \cite{Eremeev.jac2017} in the case of the single SL, while for the thicker films and bulk
it means the \emph{inter}layer AFM state (Fig.~\ref{fig1}c). cAFM (uAFM) stands for the compensated 
(uncompensated) AFM state. 
$T_c$ denotes the Curie or N\'eel temperature in the FM or AFM cases, respectively.
The numbers in brackets indicate the error bar.
Details of the Monte Carlo simulations can be found in Supplementary Note I.
}
\label{tab:mag}
\begin{center}
\begin{tabular}%
{>{\PBS\centering\hspace{0pt}} p{1.50cm}%
 >{\PBS\centering\hspace{0pt}} p{2.55cm}
 >{\PBS\centering\hspace{0pt}} p{1.10cm}
 >{\PBS\centering\hspace{0pt}} p{1.50cm}
 >{\PBS\centering\hspace{0pt}} p{1.20cm}}
\hline
\hline
Thickness (SL)   & $\Delta E_\text{A/F}$ (meV/(Mn pair)) & Order         & MAE (meV/Mn)   & $T_c$ (K) \\
\hline
\hline
1                &   14.77                & FM            & 0.125       & 12(1)     \\
\hline
2                &  -1.22                 &cAFM           & 0.236       & 24.4(1)     \\
3                &  -1.63                 &uAFM           & 0.215       &           \\
4                &  -1.92                 &cAFM           & 0.210       &           \\
5                &  -2.00                 &uAFM           & 0.205       &           \\
6                &  -2.05                 &cAFM           &             &           \\
7                &  -2.09                 &uAFM           &             &           \\
\hline
$\propto$ (bulk) & -2.80                  &cAFM           & 0.225       & 25.42(1)   \\
\hline
\hline
\end{tabular}
\end{center}
\end{table}
}

For the 2-SL-thick film, the total-energy calculations show that the \emph{inter}layer 
coupling is AFM, leading to the compensated AFM (cAFM; Fig.~\ref{fig1}c) ordering as in the bulk material 
\cite{Eremeev.jac2017, Otrokov.arxiv2018}. Thickness increase up to 3 SLs keeps the 
interlayer exchange coupling antiferromagnetic, but, since the number of blocks is odd, 
an uncompensated AFM (uAFM; Fig.~\ref{fig1}c) state arises. 
Similarly to the 2-SL- and 3-SL-thick films cases, we also predict cAFM and uAFM states for the thicker films made of even and odd number of SLs, respectively (Table~\ref{tab:mag}). 

It is well known that the magnetic anisotropy and the interlayer exchange coupling play 
a crucial role in (quasi-)2D magnets. Indeed, if a purely 2D magnet has an easy-plane 
magnetic anisotropy, it features no magnetic order at any temperature except 0~K 
according to the Mermin-Wagner theorem \cite{Mermin.prl1966, Bander.prb1988}. The reason 
for this is the Goldstone mode of the gapless long-wavelength excitations, whose 
destructive role increases with decreasing dimensionality of the system. In the limit of 
strong easy-plane anisotropy such systems, instead of a second order phase transition, 
were shown to undergo a so-called Berezinskii-Kosterlitz-Thouless 
transition \cite{Berezinskii.jetp1971, Kosterlitz.jpc:ssp1973}, which is manifested in a 
change of the spin-spin correlation function behavior from a power law below the crossover 
temperature $T_\text{BKT}$ to an exponential law above it. It is precisely the interlayer 
exchange coupling that stabilizes the long-range order at finite temperatures in such 
cases \cite{Katanin.pusp2007}. Alternatively, even a small gap in the excitation spectrum 
introduced by the easy-axis magnetic anisotropy can significantly reduce the impact of the 
low-energy excitations. In this case, the three-dimensional (3D) exchange contribution is 
expected to further enhance the critical temperature \cite{Otrokov.prb2012}. 

\begin{figure}[!bth]
\begin{center}
\includegraphics[width=0.40\textwidth]{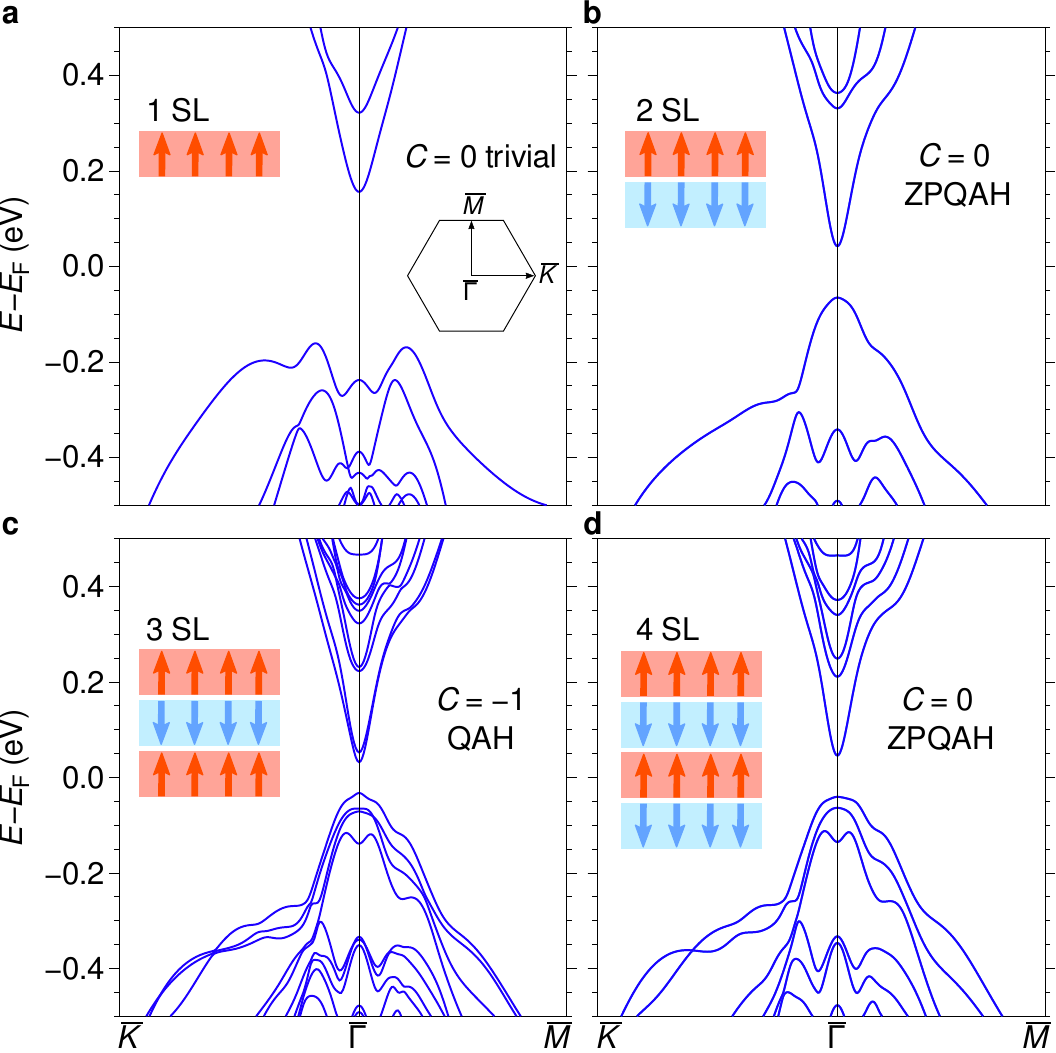}
\end{center}
\caption{ Electronic band structures of the \MBT\, films calculated
along the $\overline{K}$--$\overline\Gamma$--$\overline{M}$ path in the 2D Brillouin zone for
different thicknesses: (a) 1 SL, (b) 2 SLs, (c) 3 SLs, and (d) 4 SLs.
}
 \label{fig:topo}
\end{figure}

We then calculate the MAE for the \MBT\, films from 1 to 5 SLs as well as for bulk (Table~\ref{tab:mag}). In all these cases, MAE is positive, 
indicating an out-of-plane easy axis in agreement with recent experiments \cite{Otrokov.arxiv2018, 
Gong.arxiv2018}. The anisotropy of the SL-thick FM film turns out to be weaker than that of the 
thicker films with AFM interlayer coupling, for which the MAE was found to be close to the bulk 
value (see Table~\ref{tab:mag}). The magnitudes of the local magnetic moments are practically 
independent from the film thickness, being roughly equal to 4.6 $\mu_B$ in all cases. 
The Curie temperature of the SL-thick FM film, that represents a purely 2D magnetic system, appears 
to be approximately equal to $\SI{12}{\kelvin}$.
Due to appearance of the interlayer exchange coupling and an increase in the MAE, the N\'eel temperature of a double SL film enhances to $\approx\SI{24.4}{\kelvin}$, which 
is just slightly lower than that of bulk (Table~\ref{tab:mag}).

Now we show that in the thin-film limit not only the magnetic, but also the electronic and topological
properties of \MBT\, are strongly thickness dependent. The bandstructure of the \MBT\, single SL block
is shown in Fig.~\ref{fig:topo}a. In agreement with the experimental data \cite{Gong.arxiv2018},
it shows an indirect bandgap of $\sim\SI{0.32}{\eV}$. 
The Chern number calculations reveal a $C=0$ state, the system being a topologically trivial ferromagnet (Table~\ref{tab:topo}).

\begin{figure}[!bth]
\begin{center}
\includegraphics[width=0.40\textwidth]{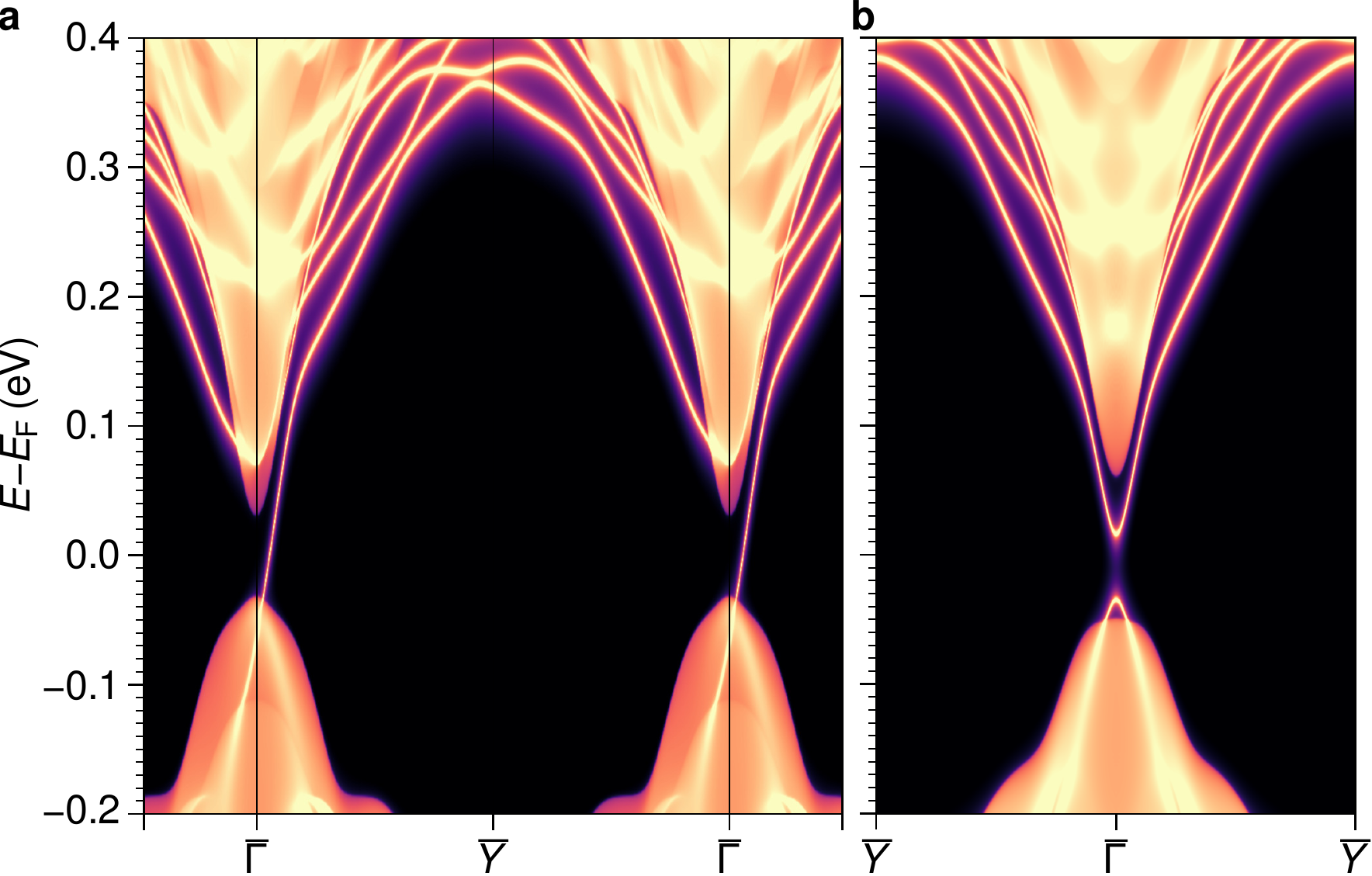}
\end{center}
\caption{Edge electronic band structures of the \MBT\, 2-SL-thick film calculated 
for the (a) FM and (b) cAFM states. The regions with a continuous spectrum correspond
to the 2D bulk states projected onto a 1D Brillouin zone. The edge crystal structure is
shown in Supplementary Note I.}
 \label{fig:rib}
\end{figure}

Upon increasing thickness up to 2 SLs the interlayer cAFM order sets in, 
leading to a doubly degenerate band spectrum (Fig.~\ref{fig:topo}b) and 
 $C=0$ again. For this system, we find a bandgap of 107 meV. 
However, if calculated in the artificial FM phase of the 2-SL-thick film, 
the Chern number appears to be equal to $-1$ indicating a QAH insulator state.
Accordingly, the edge band structure of the system shows a single 1D chiral mode 
(Fig.~\ref{fig:rib}a). Reversing the magnetization of the FM 2-SL-thick 
{\renewcommand{\arraystretch}{1.00}%
\begin{table}[!bth]
\caption{
Thickness dependence of the MnBi$_2$Te$_4$ films topology and bandgap size. QAH and ZPQAH stand for
the quantum anomalous Hall phase and its zero plateau state, respectively. 
}
\label{tab:topo}
\begin{center}
\begin{tabular}%
{>{\PBS\centering\hspace{0pt}} p{2.30cm}%
 >{\PBS\centering\hspace{0pt}} p{2.80cm}
 >{\PBS\centering\hspace{0pt}} p{2.80cm}}
\hline
\hline
Thickness (SL)    & Topology    & Bandgap (meV)\\
\hline
\hline
1                 & Trivial            & 321 \\
2                 & ZPQAH              & 107 \\
3                 & QAH                &  66 \\
4                 & ZPQAH              &  97 \\
5                 & QAH                &  77 \\
6                 & ZPQAH              &  87 \\
7                 & QAH                &  85 \\
\hline
$\propto$ (bulk)  & 3D AFM TI          &  225 \\
\hline
\hline
\end{tabular}
\end{center}
\end{table}
}
\MBT\, film yields the $C=+1$ QAH state.
These results suggest that the 2-SL-thick cAFM \MBT\, film is likely to be in a 
so-called zero plateau QAH (ZPQAH) state. Up to now, the ZPQAH state was an artificial state
of a QAH insulator that is realized in the process of the magnetization 
reversal by external magnetic field (i.e. during the transition between 
the two QAH states with Chern numbers of opposite signs). ZPQAH state manifests itself in the appearance 
of flat regions in the hysteresis-like dependence of the Hall
conductivity on the external field $\sigma_{xy} (H)$. Namely, within 
certain range of $H$ close to the coercivity, the $\sigma_{xy}=0$ 
plateau is observed, which corresponds to a fully gapped band structure. 
Outside this $H$ range, $\sigma_{xy}$ rapidly reaches a quantized 
value of either $+e^2/h$ or $-e^2/h$, depending on the magnetization
direction. Such a situation can be achieved either in (i) a zero magnetization 
state of the magnetically-doped QAH insulator \cite{Wang.prb2014} due to 
the coexistence of the upwards and downwards magnetized domains or (ii)  
the antiparallel magnetizations state of an FM1($\uparrow$)/TI/FM2($\downarrow$) QAH 
heterostructure, where FM1 and FM2 are two different FM insulators \cite{Wang.prb2015}. 
To check whether the ZPQAH state is realized in the 2-SL-thick \MBT\, film, we have calculated the edge band structure in the cAFM
ground state of the system.  We find a fully gapped spectrum
(Fig.~\ref{fig:rib}b), corresponding to $\sigma_{xy}=0$.
Thus, the 2-SL-thick \MBT\, film represents first ever example of an  
intrinsic ZPQAH phase.

At a thickness of 3 SLs, the system enters in a $C=-1$ QAH
insulator state with a bandgap of $\sim$66 meV. 
Similarly to the 2-SL-thick (3-SL-thick)
film cases, we also predict the intrinsic ZPQAH (QAH) states for
the 4- and 6-SL-thick (5- and 7-SL-thick) films, respectively
(Table~\ref{tab:topo}). 

At this point, having described various phases 
realized in the MnBi$_2$Te$_4$ films, it is important to stress a crucial advantage of the here proposed
(ZP)QAH insulators: they show ordered structures, inherent to the stoichiometric material, which guarantees
them against disorder-related drawbacks such as the bandgap fluctuation \cite{Lee.pnas2015} or
superparamagnetism \cite{Lachman.sciadv2015, Krieger.prb2017}. This very fact, together with the large
bandgaps of such systems (Table~\ref{tab:topo}), could facilitate the observation of the QAHE at temperatures notably higher than those
achieved so far. This is all the more true since, from the 2 SLs thickness, the MAE is already close to that of the bulk, indicating that the \MBT-based QAH insulators
should have critical temperatures comparable to the bulk N\'eel temperature of \MBT\, (Table~\ref{tab:mag}).
To be mentioned as well is a solid state realization of axion electrodynamics in a ZPQAH state proposed recently \cite{Wang.prb2015}.
Up to now the axion insulator state was being sought for in the FM1/TI/FM2 QAH heterostructures.
In such systems, a relatively thick TI spacer enables magnetization reversal of the individual FM layers that have different coercivities, leading to the overall AFM alignment and,
consequently, to a ZPQAH state \cite{Mogi.sciadv2017, Xiao.prl2018}. In contrast to the latter heterostructures, 
the \MBT\, thin films made of even number of SLs realize this state intrinsically,  i.e. 
 without need of magnetic field application.

In summary, using \emph{ab initio} and Monte Carlo calculations, we have scrutinized the magnetic, electronic
and topological properties of the \MBT\, AFM TI thin films. Belonging
to the class of layered vdW compounds, in the 2D limit \MBT\, shows a 
unique set of thickness-dependent transitions through various 
phases, being among them wide-bandgap QAH and ZPQAH states. Similar 
behaviour can possibly take place in other compounds of 
the \MBT\, family, such as \MST, \MBS, and others. We believe that
our findings will stimulate intensive studies of thin films of vdW antiferromagnets as
prospective materials for AFM spintronics.

\section*{Acknowledgments}
Authors thank J.I. Cerd\'a and V.N. Men'shov for stimulating discussions.
We acknowledge the support by the Basque Departamento de Educacion,
UPV/EHU (Grant No. IT-756-13), Spanish Ministerio de Economia y
Competitividad (MINECO Grant No. FIS2016-75862-P), Academic
D.I. Mendeleev Fund Program of Tomsk State University (Project No.
8.1.01.2018), and Fundamental Research Program of the State Academies of Sciences for 2013--2020, line of research III.23.
The support by the Saint Petersburg State University
for scientific investigations (Grant No. 15.61.202.2015) is also
acknowledged. I.P.R. acknowledges support by the Ministry of 
Education and Science of the Russian Federation within the 
framework of the governmental program ‘‘Megagrants’’ (state task 
no. 3.8895.2017/P220). This study was supported by Russian Science 
Foundation No. 18-12-00169 {}in the part of the calculations within 
tight-binding method. A.E. acknowledges financial support from DFG 
through priority program SPP1666 (Topological Insulators). The calculations were performed in Donostia
International Physics Center, in the Research park of St. Petersburg
State University Computing Center
(http://cc.spbu.ru), and at SKIF Cyberia cluster of Tomsk State University. \\


\end{document}